\documentclass[aip,rsi,showkeys,preprint]{revtex4}

\usepackage{graphicx}
\usepackage{dcolumn}
\usepackage{bm}
\usepackage{subfigure}

\begin{document}

\title{Nanoradian angular stabilization of x-ray optical components}

\author{Stanislav Stoupin}
\email{sstoupin@aps.anl.gov}
\author{Frank Lenkszus}
\author{Robert Laird}
\author{Kurt Goetze}
\author{Kwang-Je Kim}
\author{Yuri Shvyd'ko}
\affiliation{Advanced Photon Source, Argonne National Laboratory, Illinois, 60439, USA}

\date{\today}

\begin{abstract}
An x-ray free electron laser oscillator (XFELO) has been recently proposed [K. Kim, Y. Shvyd'ko, and S. Reiche, Phys. Rev. Lett. 100, 244802 (2008)]. Angular orientation and position in space of Bragg mirrors of the XFELO optical cavity must be continuously adjusted to compensate instabilities and maximize the output intensity. An angular stability of about 10 nrad (rms) is required [K. Kim and Y. Shvyd'ko Phys. Rev. STAB 12, 030703 (2009)]. To approach this goal, a feedback loop based on a null-detection principle was designed and used for stabilization of a high energy resolution x-ray monochromator ($\Delta E/E \simeq 4 \times 10^{-8}$, $E$ = 23.7 keV) and a high heat load monochromator. Angular stability of about 13 nrad (rms) has been demonstrated for x-ray optical elements of the monochromators.
\end{abstract}

\maketitle
\section{Introduction}

Implementation of an x-ray optical cavity plays a crucial role in research efforts towards the realization of an x-ray free electron laser oscillator (XFELO) \cite{Kim08,Kim09}. Stable operation of an XFELO can only be achieved through precise control of the cavity geometry. An angular stability of 10 nrad (rms) is required to generate coherent monochromatic radiation using successive Bragg reflections from cavity crystals \cite{Kim09}. A variety of technical solutions can be considered. Among them, of major practical importance is stabilization by direct feedback on the signal of interest, the XFELO output, as opposed to indirect stabilization (e.g., using a visible light optics scheme with reflectors coupled to the Bragg mirrors of the cavity). 
An additional requirement is to stabilize many optical axes by using multiple feedback channels operated with only one common detector on the signal of interest. Such method would allow to substantially reduce the number of x-ray detectors involved in operation of the XFELO.

A number of factors are expected to destabilize optical elements of the cavity including varying heat load, mechanical instabilities, ambient conditions, etc. In general, any factor known to affect properties of diffracting crystals for high-energy-resolution applications will contribute to destabilization. Angular positions of diffracting crystals must be continuously adjusted to compensate for the instabilities and maximize the output intensity of a monochromatic x-ray source. To approach the desired angular stability requirement, a feedback scheme known as automated control of Bragg angle \cite{Mellaert70} was adapted 
in the present work for an x-ray monochromator with
relative energy resolution of $\Delta E/E \simeq 4 \times 10^{-8}$ ($E \simeq$ 23.7~keV) at the inelastic x-ray scattering beamline XOR/IXS 30-ID of the Advanced Photon Source. The angular region of the monochromator stability is approximately 50 nrad and is therefore appropriate for these feasibility studies. The feedback signal is proportional to the slope of the reflection curve of the monochromator crystals. The signal is extracted from intensity of x-rays reflected from the monochromator crystals using a so-called null-detection principle. Previously, this technique has been widely used for stabilization of double crystal x-ray monochromators with typical energy resolution of about 1 eV \cite{Mills80,Bridges87,Ramanathan88,Fischetti04,Proux06}. 
An angular stability in the $\mu$rad regime is required for such monochromators due to a few $\mu$rad angular widths $\delta \Theta$ of the commonly used low index Bragg reflections (e.g., for Si(111) 
and C(111) $\delta \Theta \gtrsim 10 \mu$rad for $E \lesssim$ 23.7~keV).

In this work we show that the same method is applicable to elements of x-ray optics with an energy resolution $\Delta E \simeq$ 1~meV. Angular stability of $\approx$ 13 nrad (rms) was demonstrated. In particular, performance of the high-resolution optical components was improved beyond the stability feasible by their mechanical design \cite{Shu00,Shu01}. Automated control of angular positions is demonstrated for optical elements of a double crystal high heat load monochromator (HHLM) and those of the high-energy-resolution monochromator (HRM). It is shown that the automated control ensures stable operation of the two monochromators which substantially improves performance of the beamline. The chosen approach can be demonstratively extended to include individual feedback channels for every diffracting element in a multi-component system while extracting the individual feedback signals from the common output signal of the system \cite{Whitcomb_pc}.

\section{Principle of operation}\label{principle}

\begin{figure}
\centering
\includegraphics[scale=0.45]{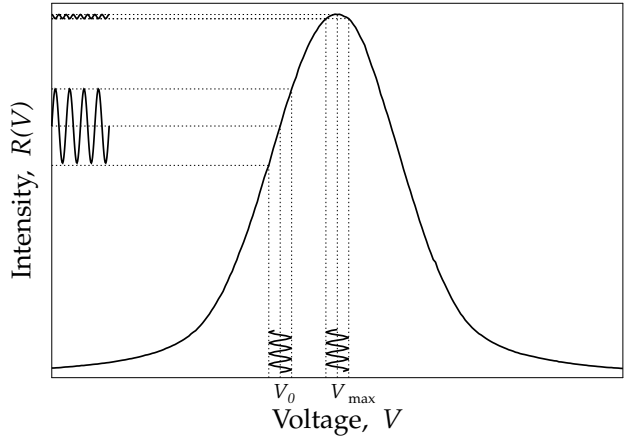}
\caption{Intensity of the reflected x-rays as a function of a voltage applied to the piezo actuator (i.e., reflection curve of the crystal). If the oscillating reference signal is applied with an onset voltage $V_0$ on the slope of the reflection curve, the amplitude of the response is large compared to the case when the onset voltage is near the position of the maximum ($V_{\rm max}$).}
\label{fig:principle}
\end{figure}

The principle of operation of the feedback is based on the fact that small deviations from 
an optimal state of a system produce negligible variations of the output signal of interest
on one hand, and, on the other hand, much larger signal variations if the system is far from that state. This principle is illustrated in Figure~\ref{fig:principle}. 
An angular position of a diffracting crystal is modulated using a signal from a reference oscillator. The oscillator signal drives a piezo actuator which controls the crystal's angular position. 
In Figure~\ref{fig:principle} the intensity of the reflected x-rays $R(V)$ is plotted vs. voltage $V$ applied to the piezo actuator (i.e., reflection curve of the crystal).
The amplitude of the resulting variation in the intensity depends on a current angular position on the reflection curve defined by an onset voltage $V_0$.
If the onset voltage is on the slope of the reflection curve, the amplitude of the response is large compared to the case when the onset voltage is near the position of the maximum ($V_{\rm max}$). Thus, the system acts as a nonlinear transformation of the reference signal.

In the following consideration phase delay in the system is neglected for illustrative purposes. It is assumed that all components of the system immediately respond to the oscillating voltage. The total voltage on the piezo actuator is $V(t)=V_0+v(t)$.
If the amplitude $v_0$ of the reference signal $v(t)=v_0 \sin(\omega t + \phi)$ is sufficiently small (i.e., much smaller than FWHM of the reflection curve) the response $R(V(t))$ can be represented using Taylor expansion in $v(t)$: 

\begin{equation}
R(V(t)) \simeq R(V_0)+\frac{dR}{dV}(V_0)v(t)+ O(v(t)^2)
\label{eq:R}  
\end{equation}
The component of the above expansion containing the first derivative (linear response) oscillates with the same reference frequency $\omega$ 
while the other components are either frequency independent or higher harmonics of $\omega$. 
For values of $V_0$ on the slope of the reflection curve, the linear response is maximized. When $V_0$ approaches $V_{\rm max}$ the linear response approaches zero.

For example, the first derivative of a quadratic profile 
$R(V)=R_{\mathrm{max}}-B(V-V_{\mathrm{max}})^2$ is :

\begin{equation}
 \frac{dR}{dV}(V)=-2B(V-V_{\mathrm{max}})
\label{eq:R2}  
\end{equation}
In this case, the linear response represented by Eq.~\ref{eq:R2} is proportional to a deviation of voltage $V$ from its optimal value $V_{\rm max}$ which maximizes the reflected intensity. 

Lock-in detection with the reference oscillator frequency allows to extract the derivative of the reflection curve from $R(V(t))$ to form a correction signal $\Delta V$ to $V_0$. It is achieved by  multiplication $R(V(t)) v(t)$ and a consequent filtering using a low pass filter with a cut-off frequency well below $f = 2\pi/\omega$. The cut-off frequency defines the response time of the feedback loop. 
To close the feedback loop the correction signal taken with an appropriate polarity is added to the voltage that drives the piezo actuator. Thus, the combined signal on the piezo actuator becomes $V(t)=V_0+\Delta V+v(t)$. Here, the correction signal $\Delta V$ is automatically optimized to minimize the linear response of the system (i.e., null-detection method) while the output x-ray intensity is maximized. If destabilizing factors cause a drift of the reflection curve to a new optimal voltage, the feedback loop generates a new correction signal through minimization of the linear response. 
Thus, losses in the output intensity due to destabilizing factors are compensated.

\section{Implementation}

\subsection{General description of the feedback system}
\begin{figure}
\centering
\includegraphics[scale=0.45]{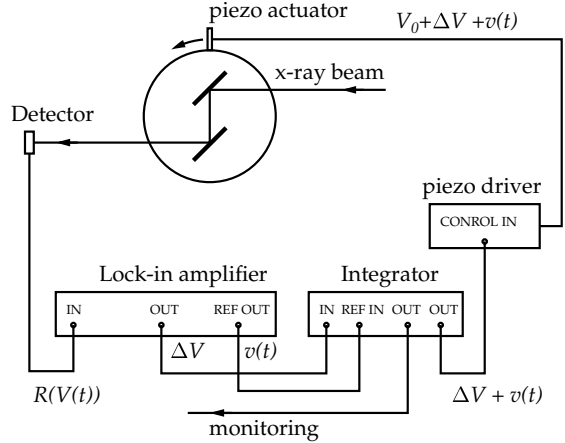}
\caption{Block diagram of the feedback loop.}
\label{fig:scheme}
\end{figure}

The null-detection feedback scheme is shown in 
Figure~\ref{fig:scheme}. The setup includes: x-ray optical component (a pair of diffracting crystals in non-dispersive configuration), a piezo actuator which performs continuous angular motion of the optical component, a detector of the transmitted intensity, a lock-in amplifier, an integrator device, and a piezo driver. A lock-in amplifier generates the oscillating reference signal which is added to the correction signal by the integrator device. The resulting signal is sent to a control input of the piezo driver. The oscillatory motion of the piezo actuator produces modulation of the reflected x-ray intensity which is demodulated by the lock-in amplifier to form the correction signal. The integrator has an additional output for monitoring variations in the correction signal. These variations provide information on the drift of angular position of the reflection curve maximum.

\begin{figure*}
\centering
\includegraphics[scale=0.9]{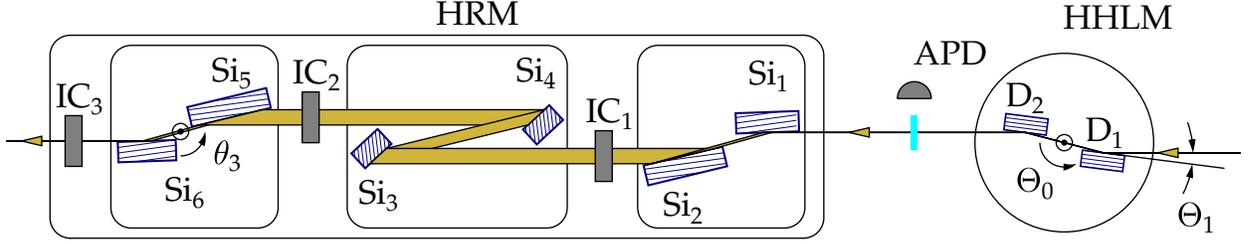}
\caption{Successive application of the two monochromators at the inelastic x-ray scattering beamline XOR/IXS 30-ID. Preliminary monochromatization is performed using the high-heat-load monochromator (HHLM) with energy bandwidth of $\Delta E_1 \simeq$ 1-2~eV. The high-resolution monochromator (HRM) selects x-ray photons to the desired bandwidth of $\Delta E_2 \simeq$ 1~meV \cite{Toellner09}.
The HHLM is a double crystal diamond (1 1 1) monochromator (crystals $\rm D1$ and $\rm D_2$).
Energy selection by HHLM is performed by variation of $\Theta_0$ 
(rotation of the two crystals around a common axis).
The high resolution monochromator consists of 3 pairs of Si crystals: 
$\rm Si_1,Si_2$ and $\rm Si_5,Si_6$ are asymmetric crystals coupled by a weak-link mechanism \cite{Shu01} 
(the first pair and the third pair respectively); 
crystals $\rm Si_3$ and $\rm Si_4$ form an asymmetric channel-cut which is cooled using liquid nitrogen.
The feedback has been introduced for the angular position $\Theta_1$ of crystal $\rm D_1$ 
and for the angle of the third pair $\theta_3$ (i.e., angle of rotation of the linked 
$\rm Si_5$ and $\rm Si_6$ crystals around a common axis).
The intensity of x-rays transmitted through the HHLM is measured with an 
avalanche photodiode (APD). The intensity of x-rays after each crystal pair 
of the HRM is monitored using ion chambers ($\rm IC_1, IC_2, IC_3$).}
\label{fig:hrm_and_hhlm}
\end{figure*}

Monochromatization of x-rays at the inelastic x-ray scattering beamline XOR/IXS 30-ID 
is achieved by successive applications of the two x-ray monochromators as
shown in Figure~\ref{fig:hrm_and_hhlm}. Preliminary monochromatization is achieved by a double crystal high-heat-load monochromator (HHLM) which selects photons to a bandwidth of $\Delta E_1 \simeq$ 1-2~eV. 
In the next step, the high-resolution monochromator (HRM) selects photons within a desired bandwidth of $\Delta E_2 \simeq$ 1~meV \cite{Toellner09}.
Two similar feedback channels have been implemented to provide automated control for
the angle $\Theta_1$ of the first crystal $\rm D_1$ of the double crystal HHLM and the angle $\theta_3$ of the third crystal pair of the HRM (i.e., angle of rotation of crystals $\rm Si_5$ and $\rm Si_6$ in the diffraction plane around their common axis - see Figure~\ref{fig:hrm_and_hhlm} for details). The common components of the two feedback loops are a commercial lock-in amplifier and an in-house built integrator. Technical details on these two are given in Section \ref{SR&Int}. Specific details about each feedback channel pertaining to HHLM and HRM are presented in Section \ref{HHLM} and Section \ref{HRM} respectively.

\subsection{Lock-in amplifier and Integrator}\label{SR&Int}

Stanford Research lock-in amplifiers (model: SR830) were used in both feedback channels.
The chosen reference oscillator parameters, amplitude and frequency, are specific for each channel and reported in Section \ref{perf}. The lock-in amplifiers were remotely controlled via the RS-232 interface. The remote control for the amplifiers has been implemented within a beamline control software EPICS.

\begin{figure*}
\centering
\includegraphics[scale=0.8]{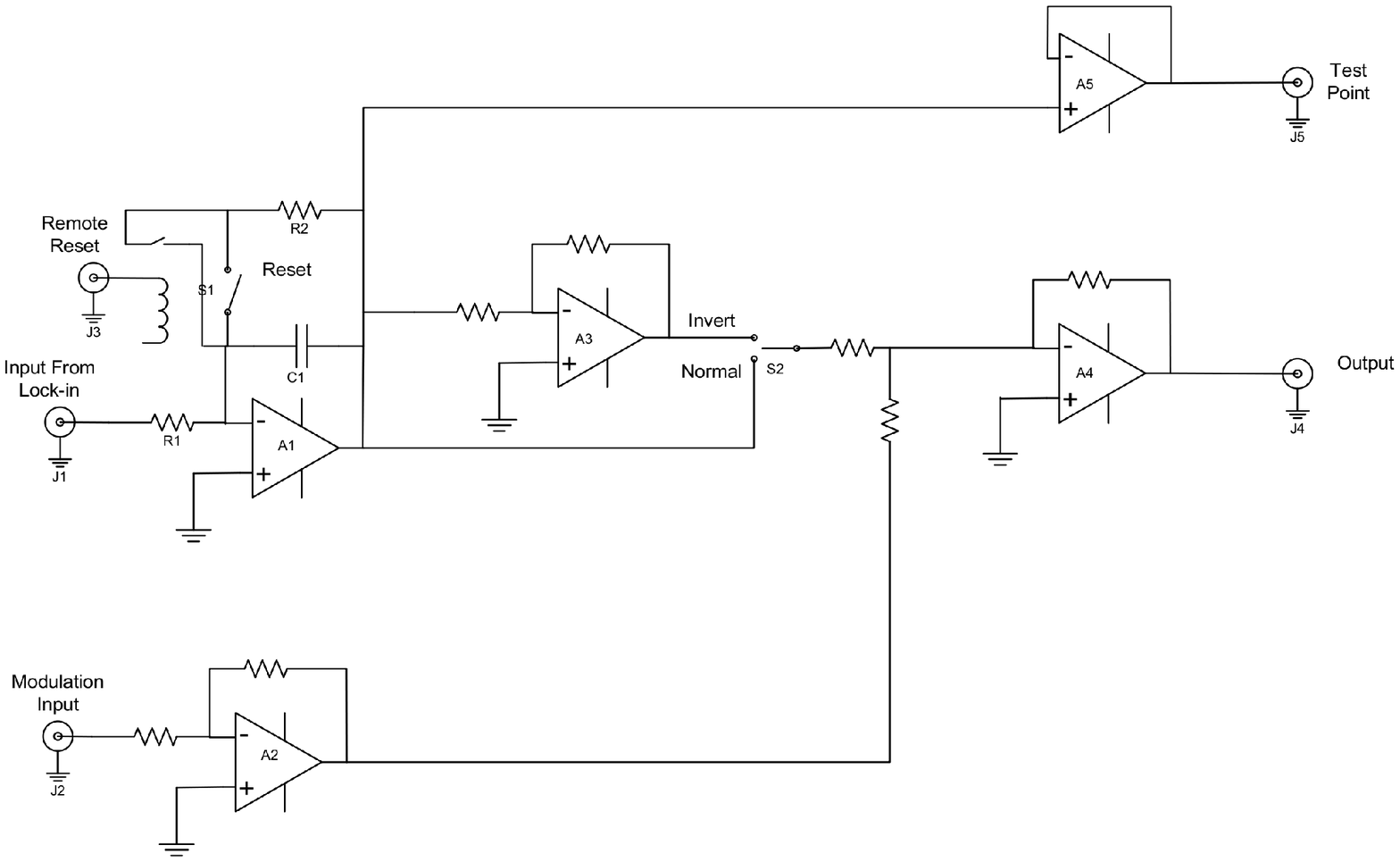}
\caption{Schematic diagram of the integrator.}
\label{fig:Int}
\end{figure*}

The integrator was built using an analog circuitry as shown in Figure~\ref{fig:Int}.
Operational Amplifier $A_1$ which forms the integrator, is an ultra-low bias current Difet device. It features a typical bias current of 30 fA. This low bias current helps to minimize static errors.  
The idealized transfer function of this integrator circuit formed by $A_1$ is:
\begin{equation}
K(s) = \frac{V_{out}}{V_{in}} = \frac{1}{sR_1C_1}
\label{eq:Int}  
\end{equation}
$C_1$ is a low leakage polypropylene type capacitor. The integrator gain which is determined by the $R_1C_1$ value is set to 1 second. The gain can be modified by changing either $R_1$ or $C_1$. The integrator may be reset either locally with switch $S_1$ or remotely via the Remote Reset input. The Remote Reset input operates a relay to discharge the integrator capacitor. In either case, the capacitor is discharged through resistor $R_2$. Note that the discharge time constant is determined by $R_2C_1$ so the reset must be held for a sufficient duration to discharge the capacitor. 

The output of the integrator may be inverted by setting switch $S_2$ to "Invert". This allows the feedback sign to easily be set properly for negative feedback with different detector/actuator configurations. The modulation input is buffered via amplifier $A_2$ and summed with the integrator value at amplifier $A_4$. The combined integrator/modulation signal is buffered via $A_4$ and made available on the "Output" connector, $J_4$. This output is capable of driving $\pm$ 10 Volts at 40 mA. The Test Point output provides means to monitor the integrator exclusive of the modulation.
%

\subsection{High Heat Load Monochromator} \label{HHLM}
The common angle of rotation $\Theta_0$ of the crystals $\rm D_1$ and $\rm D_2$ defines the energy of the x-rays transmitted through the HHLM (Figure~\ref{fig:hrm_and_hhlm}) which is typical for x-ray double crystal pre-monochromators. To optimize the transmitted intensity the angular position of the first crystal $\Theta_1$ has to be adjusted such that the reflecting planes of $\rm D_1$ and $\rm D_2$ are 
parallel. The angle $\Theta_1$ is adjusted independent of the angle $\Theta_0$. 

The transmitted intensity is monitored using an avalanche photodiode (APD) which detects photons scattered from a low absorption beryllium film placed in the vacuum flight-path of the x-ray beam.
The output pulses from the APD are processed using a fast leading edge discriminator.
The NIM output signal of the discriminator is converted to an analog signal using an RC filter with a bandwidth of 260 Hz. The resulting analog signal is employed for the HHLM feedback.

The angular motion of the first crystal is performed by a high precision piezo actuator (DPT-C-S-VAC, Queensgate) with an embedded position sensing circuity. The actuator's piezo driver (AX101, Queensgate) involves a servo loop which is used to reduce non-linearity of the actuator to achieve high repeatability of its motion. The piezo driver has a separate analog input (control input in Figure~\ref{fig:scheme}) which accepts external signals to be combined with the onset voltage on the actuator.

The reflection curve of the first crystal was measured by step-scanning a voltage on the piezo actuator through a range that included the reflection curve peak and its tails. The curve is
shown in Figure~\ref{fig:principle}. The width of the reflection curve $\rm FWHM \simeq 0.67$ V was determined from a Gaussian fit. 

In order to determine the range of frequencies where the reference signal is effectively converted into oscillations in x-ray intensity, a dynamic response of the open feedback loop was measured. Detailed description and results of the measurements are given in Section \ref{TF}.
%
\subsection{High Resolution Monochromator} \label{HRM}
The HRM is a part of high-energy-resolution inelastic x-ray scattering spectrometer (HERIX). 
The monochromator designed by T. Toellner \cite{Toellner09} operates in the energy range of 
23.7-29.7~keV. It involves three pairs of diffracting crystals as shown in Figure \ref{fig:hrm_and_hhlm}. The first pair ($\rm Si_1$ and $\rm Si_2$) consists of two asymmetric Si crystals using low-index Bragg reflections. The two crystals are coupled by a weak-link mechanism \cite{Shu01}. This pair is used to reduce the angular divergence of the x-ray beam to $\simeq$ 0.35 $\mu$rad which is crucial for the monochromatization. The second pair ($\rm Si_3$ and $\rm Si_4$) is a liquid nitrogen cooled Si channel-cut
\cite{Toellner05}
using high-index Bragg reflections and is the actual monochromator.
The third crystal pair ($\rm Si_5$ and $\rm Si_6$) is similar to the first pair and is used to restore the size of the beam to the original size of the incident pre-monochromatized beam.
The angular acceptance of the third pair is comparable to the angular divergence $\simeq$ 0.35 $\mu$rad of the incident beam (i.e., the x-ray beam after the second pair). The x-ray intensity is transmitted through the third pair only within a very small angular range. Therefore, disturbances in the angular position of the third pair cause one of the main sources of the instability. 
The intensity of transmitted x-rays is monitored after each pair using ion chambers ($\rm IC_1, IC_2, IC_3$). A signal from $\rm IC_3$ represents the intensity of the transmitted beam with 1~meV bandwidth and is used in the HRM feedback loop. 

\begin{figure}[h]
\centering
\includegraphics[scale=0.48]{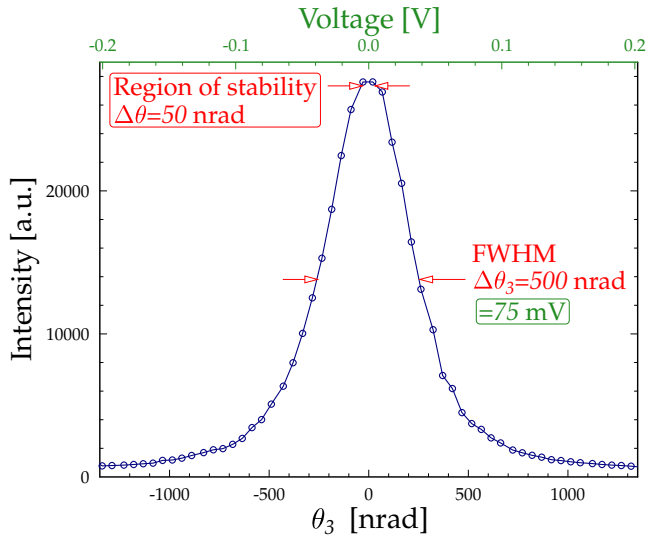}
\caption{Reflection curve of the third crystal pair (HRM) measured in units of angle and voltage. The required region of angular stability is about 50 nrad.}
\label{fig:rcurve}
\end{figure}

Rotation of the third crystal pair ($\theta_3$) is performed by a high precision stage KTG-15D 
(Kohzu). The same rotational motion is coupled to a piezo actuator P-753.11C (Physik Instrumente). The actuator is powered by a piezo driver module E-503 (Physik Instrumente). The onset voltage on the actuator can be controlled remotely using beamline software. These two options for motion control allow independent measurements of the reflection curve of the third pair in units of angle or voltage. 
The measured reflection curve of the third pair is shown in Figure \ref{fig:rcurve} where the x axis 
on the bottom is in angular units (nrad) and the one on the top is in units of voltage. 
The measured FWHM is about 500 nrad or equivalent 0.76 V (both evaluated from Gaussian fits). These give a conversion factor $\eta_0$ = FWHM($\mu$rad)/FWHM(V) = 0.7 $\mu$rad/V between the angle and the voltage scales. The piezo driver module has a constant gain factor of 10. The conversion factor determined above becomes $\gamma_0$ = 7 $\mu$rad/V with respect to the correction signal applied 
to the control input of the module (i.e., the integrator output). 

A reasonable stability requirement is that the angle of the third pair should be within 1/10 of FWHM from the unstable angular position of the maximum (i.e., 50 nrad). In Section \ref{perf} it is demonstrated that this requirement has been satisfied using the null-detection feedback scheme. 

%
\section{Dynamic response}\label{TF}
Time delay of the feedback system has been neglected in Section~\ref{principle} to demonstrate the principle of operation. In reality, the components of the feedback loop have non-zero 
response times. This results in a phase shift between the oscillations in the x-ray intensity and those of the reference oscillator. The phase shift affects the value of the correction
signal which can be optimized by adjusting the phase of the reference oscillator. To ensure a stable phase shift in the feedback loop an off-resonance frequency of the reference oscillator has to be chosen. Therefore, knowledge of the frequency-dependent response (or dynamic response) of the feedback loop is crucial for optimization of the feedback performance. 

\begin{figure}
\includegraphics[scale=0.45]{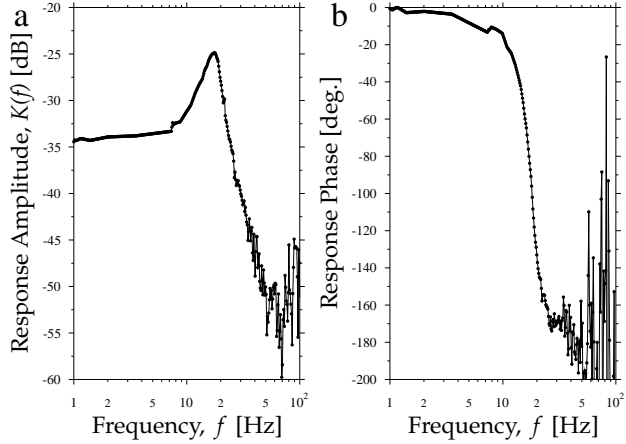}
\caption{Measured dynamic response of the HHLM feedback channel: amplitude (a) and phase (b).}
\label{fig:dynr_hhl}
\end{figure}

The dynamic response of the feedback channels was measured with an Agilent 35670A dynamic signal analyzer operated in swept sine mode. The onset voltage was set to position the operating point on the steep slope portion of the reflection curve to maximize the linearity of the response. 
An excitation level of 100~mV peak-to-peak  was applied to measure the dynamic response 
of the HHLM feedback channel. The response is of a second order type as shown in Figure \ref{fig:dynr_hhl}. The peak in the response is at approximately 16.8~Hz with a phase of 
102.3~degrees. Note that above approximately 40~Hz the measured response becomes very noisy which indicates that the received signal is below the sensitivity level of the analyzer. The observed peak is due to the first mechanical resonance of the system. Reliable operation of 
the HHLM feedback loop was achieved using an off-resonance excitation with the reference
oscillator frequency of 2~Hz.

\begin{figure}
\includegraphics[scale=0.45]{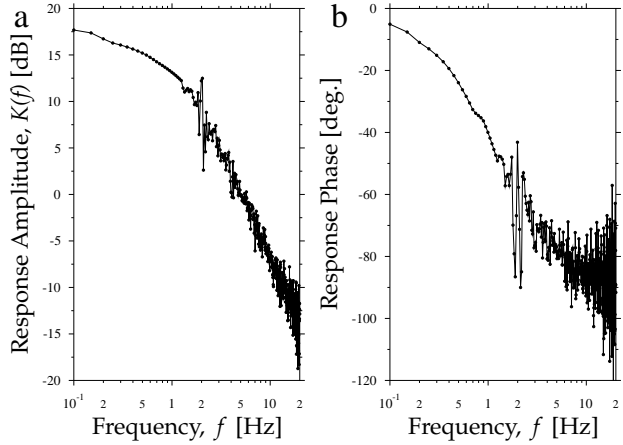}
\caption{Measured dynamic response of the HRM feedback channel: amplitude (a) and phase (b).}
\label{fig:dynr_hrm}
\end{figure}

The same strategy was applied to measure the dynamic response of the HRM feedback channel. The input level of the swept sine wave was 10~mV peak-to-peak in order to preserve linear relationship between the input and the output on the steep slope of the HRM reflection curve. The measured response rapidly decays with frequency and becomes noisy (Figure~\ref{fig:dynr_hrm}). Such behavior is due to absence of mechanical resonances at the low frequencies for the structure that supports the high resolution optics and a rapidly decaying off-resonance response of the piezo actuator \cite{Shu_pc}.

Reliable operation of the HRM feedback loop was obtained with a reference oscillator frequency of 10~Hz. The amplitude of the reference oscillator was only $v_0$ = 4~mV (rms). The measured dynamic response allows to estimate an amplitude $\theta_0$ of an angular oscillations excited by the reference oscillator. The static conversion factor $\gamma_0$ was defined in section \ref{HRM}. 
A dynamic conversion factor is obtained as: 
\begin{equation}
\gamma(f)=\gamma_0 K(f)/K(0)
\label{eq:gamma_f}  
\end{equation}
where $K(f)$ is the amplitude frequency response (Figure~\ref{fig:dynr_hrm}a). The response at 10 Hz is at a level of about -26~dB with respect to the value at the lowest frequency. This reduction results in a very small amplitude of the angular oscillations at 10~Hz: $\theta_0 = \gamma(f) v_0 \approx$ 1 nrad (rms). Thus, the angular oscillations due to the reference excitation $v(t)$ only negligibly disturb the angular position of the third pair. The angular stability is determined by the jitter in the correction signal $\Delta V$. An estimate for this limit will be given below.

\section{Performance and angular stability}\label{perf}

Both feedback loops were tested and further employed during user operations
at the beamline. The use of the HHLM feedback loop eliminates necessity of
manual correction of the angular position of the first crystal under variable heat load conditions. 
This was found especially helpful to resume experiments after storage ring beam dumps.
The operational parameters of the two feedback loops are summarized in Table \ref{tab:param}.

\begin{table}[h]
 \centering                    
 \caption{Feedback loop operating parameters} 
 \begin{tabular}{c c c}          
 \hline\hline
 Parameter & HHLM & HRM \\ [0.5ex]
 \hline 
 reflection curve FWHM, $\mu$rad & 9 & 0.5 \\
 Ref. oscillator frequency, Hz & 2 & 10 \\
 Ref. oscillator voltage, mV (rms) & 10 & 4 \\
 \hline
 \end{tabular}
 \label{tab:param}
 \end{table}
 
 \begin{figure*}[t]
\centering
\includegraphics[scale=0.9]{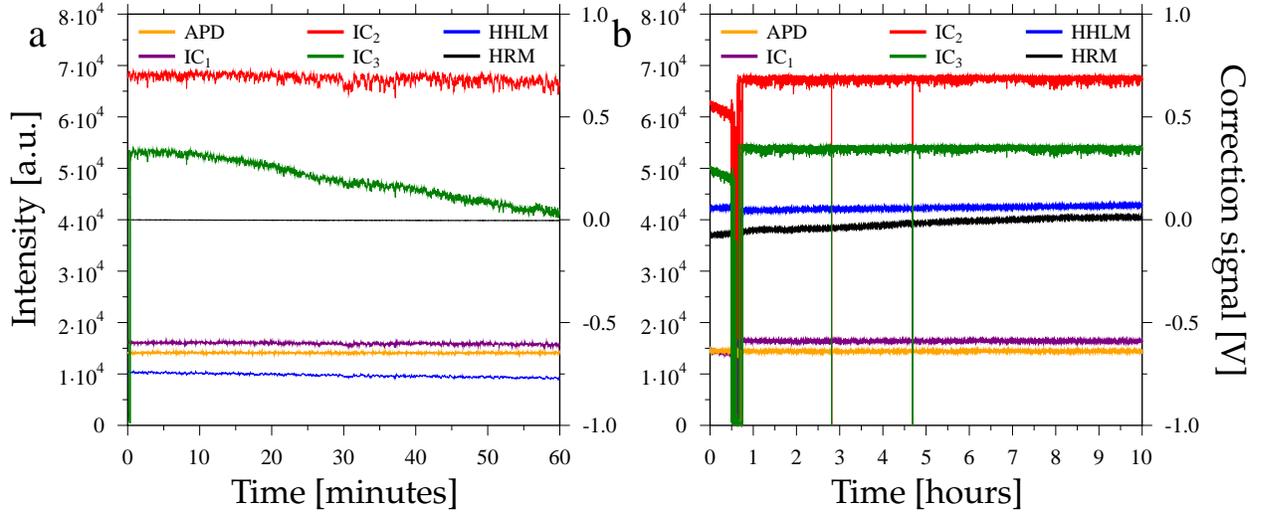}
\caption{Variation of x-ray intensities and feedback correction signals with time: without the use of feedback on the third pair of HRM (a) and with the feedback (b). The incident intensity (APD) and intensities after the first ($\rm IC_1$) and the second ($\rm IC_2$) crystal pairs of HRM remain constant. The output intensity ($\rm IC_3$) is kept constant by the feedback. The correction signals applied to the piezo drivers of the high heat load monochromator (HHLM) and the high-resolution monochromator (HRM) vary slowly to provide stabilization.}
\label{fig:fdb_mon}
\end{figure*}

The improvement in performance of the HRM resulting from the use of the null-detection
feedback is illustrated in Figure \ref{fig:fdb_mon} which shows variation in intensity of 
x-rays with time recorded using intensity monitors along the path of the x-ray beam. 
The intensity of x-rays incident to the first crystal pair of the HRM (i.e., the output 
intensity of the HHLM - orange line) is kept constant by the HHLM feedback loop. The 
correction signal of the HHLM integrator (blue line) varies slowly with time to provide 
compensation to the first crystal of HHLM. The correction signal for the angular position of 
the third pair of the HRM is shown by the black line. 
The purple line and the red line represent x-ray intensities recorded by ion chambers 
$\rm IC_1$ and $\rm IC_2$ after the first and the second crystal pairs respectively. 
These signals do not change significantly with time which suggests that initial alignment of the crystals in the first and the second pairs is preserved. Also, this means that the spectral content of the x-rays from HHLM remains stable and contains the x-ray energies selected by the first pair of the HRM (i.e., the energy drift of the HHLM is minimized). Such conditions are typical during data collection with HERIX spectrometer. 

Figure \ref{fig:fdb_mon}a represents operation of HRM under the above conditions, without 
the feedback on the third pair. Initially the third pair was aligned to provide maximum 
output intensity (green line). One hour of operation after the initial alignment yields 
20\% reduction in the output intensity as a result of naturally occurring angular 
destabilization. In contrast, continuous operation of the feedback loop allows to maintain the 
optimal level of the output intensity for many hours as shown in Figure \ref{fig:fdb_mon}b. 
Overall, the HRM feedback ensures stable operation of the monochromator and the HERIX 
spectrometer on the whole. The feedback operation was found especially helpful 
during unattended long-term data collection. In addition, the output intensity was found 
to remain unaffected during maintenance procedures such as a refill of liquid nitrogen 
for the second pair of the HRM. Prior to introduction of the feedback, 
this necessary procedure required a waiting time of a few hours to stabilize the output intensity.

The angular stability of the third pair was estimated as follows. 
The long-term variation of the HRM correction signal that yields constant output intensity (Figure \ref{fig:fdb_mon}b) was approximated with a 4th order polynomial function. 
The short-term variation $\Delta v$ of this signal was evaluated as standard deviation from the polynomial approximation which was found to be $\Delta v \approx$ 2~mV. This standard 
deviation is a measure of angular stability with characteristic time of about the time 
constant of the integrator (1~s). The upper estimate for angular stability is given by: 
$\Delta \theta = \gamma_0 \Delta v \approx$ 13 nrad (rms).

\section{Conclusions}
In conclusion, a feedback stabilization system based on null-detection principle was successfully implemented for x-ray monochromators at the 
inelastic x-ray scattering beamline XOR/IXS 30-ID of the Advanced Photon Source.
We have demonstrated angular stability of the optical elements on the level of $\approx$ 13 nrad. 
This number is very encouraging for the realization of the XFELO. The chosen method complies with the XFELO design requirements. Further improvement in the stability is anticipated. In addition, the performance of the beamline has been substantially improved as a result of added stability. 

The technical details provided in this paper may serve as a guide to improve operation of any high-resolution x-ray optical component. We note that automatic adjustment of the Bragg angle is a hardware based solution which requires a built-in piezo actuator to perform continuous angular motion of the optical element and related electronics. The feedback loop described here involves a commercially available lock-in amplifier and an integrator unit which can be easily built. In cases when the hardware solution is not applicable, an alternative software approach can be used \cite{Jemian02}. However, the software approach is limited in that the time scale of the software feedback loop is affected by time response of a beamline control software (e.g., EPICS). In this regard, stability provided by the hardware solution is superior. 

\section{Acknowledgment}
We are indebted to S. Whitcomb for very important discussions and suggestions.
T. Toellner and  D. Shu are acknowledged for the development of the high-resolution monochromator
used in these studies. T. Roberts is acknowledged for help with implementation of the feedback system.
We would like to thank J. Kropf, M. Rivers, T. Gog and D. Shu for helpful discussions on the topic and A. Said, M. Upton and A. Cunsolo for their help with testing. Special thanks are due to J. Kropf for loaning a lock-in amplifier at an early stage of the project. Use of the Advanced Photon Source was supported by the U. S. Department of Energy, Office of Science, Office of Basic Energy Sciences, under Contract No. DE-AC02-06CH11357.


\end{document}